\documentclass[aps,prb,twocolumn,showpacs,amsmath,amssymb]{revtex4-1}
\usepackage{bookmath} % definitions and shortcuts
\usepackage{graphicx}
\usepackage{amsmath}
\usepackage{mathrsfs}
\usepackage{color}

\newcommand{\Lagr}{\mathscr{L}}

\begin{document}
%~~~~~~~~~~~~~~~~~~~~~~~~~~~~~~~~~~~~~~~~~~~~~~~~~~~~~~~~~~~~~~~~~~~~~~~~~~~~~~~%
%\title{Collective modes and topological stability of domain walls in superconductors}
%\title{Massless amplitude/Higgs mode in non-uniform condensates}
%\title{Collective modes in non-uniform condensates}
%\title{Collective modes and topological stability of non-uniform superconductors} 
\title{Bound collective modes in nonuniform superconductors} 

\author{Andrew~R.~Hammer}
\author{Anton~B.~Vorontsov}

\affiliation{Department of Physics, Montana State University, Bozeman, Montana 59717, USA}

\date{\today}

\begin{abstract}
We study dynamics of a superconducting condensate 
in the presence of a domain wall defect in the order parameter.  
We find that broken translation and reflection symmetries result in 
new collective excitations, bound to the domain wall region. 
Two additional amplitude/Higgs modes lie below the bulk pairbreaking edge $2\Delta$; 
one of them is a Goldstone mode with vanishing excitation energy. 
Spectrum of bound collective modes is related to 
the topological structure and stability of the domain wall. 
The `unbound' bulk collective modes and transverse gauge field 
mostly propagate across the domain wall, 
but the longitudinal component of the gauge field is completely reflected. 
Softening of the amplitude mode suggests reduced damping and possible route 
to its detection in geometrically confined superfluids or in superconductor-ferromagnetic 
heterostructures. 
\end{abstract} 

%74.20.De	Phenomenological theories (two-fluid, Ginzburg-Landau, etc.)
%74.81.-g	Inhomogeneous superconductors and superconducting systems, including electronic inhomogeneities

\maketitle

%~~~~~~~~~~~~~~~~~~~~~~~~~~~~~~~~~~~~~
\section{Introduction}
%\emph{Introduction.}
%~~~~~~~~~~~~~~~~~~~~~~~~~~~~~~~~~~~~~
%

Observation of the Higgs particle at LHC\cite{HiggsDiscoveryRMP2014} 
has emphasized the connection between high energy and condensed matter physics 
through collective modes. \cite{Volovik:2013ko,Pekker:2015ib}
These excitations are the normal modes of order parameter (OP) fluctuations, 
reflecting the symmetry and structure of the OP's potential landscape. 
%and thus, indirectly, the nature of interactions in superfluids. 
In a singlet isotropic superconductor with a complex order parameter 
$\Delta(\vr,t) = \psi(\vr,t)\exp[i\varphi(\vr,t)]$ 
a gapless Bogoliubov-Anderson $\varphi(\vr,t)$-phase mode
\cite{Bogoliubov1947,Bogoliubov1958,Anderson1958} 
is a result of spontaneously broken %global continuous 
$U(1)$ symmetry.\cite{Nambu1960,Goldstone:1962ty} 
Interaction with electromagnetic gauge field shifts this mode up 
to plasma frequency.\cite{Anderson1963}
Fluctuations of the other degree of freedom, $\psi(\vr,t)$, represents 
the amplitude mode, often called Higgs mode, due to the close analogy to 
its particle counterpart.\cite{LRyder} %\cite{HiggsPRL1964,*EnglertPRL1964}

Detection of the amplitude mode in condensed matter systems has been a long-stading 
challenge. The original discovery of this mode in charge-density-wave material 
NbSe$_2$ \cite{Sooryakumar:1980um,LittlewoodVarma81,*LittlewoodVarma82}
highlights the main difficulty associated with the fact that its energy is $2|\Delta|$ 
leading to its quick decay into two-particle excitations. 
This search is continuing due to its fundamental importance and 
intriguing possibility of insight into Standard Model 
from low-energy experiments.\cite{Volovik:2013ko,Anderson:2015ev} 
Recently the amplitude mode 
near a quantum critical point was investigated theoretically 
\cite{Podolsky:2011ti,*Pollet2012,*Gazit:2013wu}
and experimentally in 
neutral superfluid of cold atoms.\cite{Endres:2012fb} 
Another report of amplitude mode detection in disordered superconductors\cite{Sherman:2015eh} 
was questioned in \cite{Cea2015} due to expected strong mixing of the amplitude 
and phase modes. 

%%%%%%%%%%%%%%%%%%%%%%%%%%%%%%%%%%%%%%%%%%%%%%%%%%%%%%%%%%%%%%%%%%%%%%%%%%%%%%%%%
\begin{figure}[t]
\includegraphics[height = 0.5\linewidth,width = 1.0\linewidth]{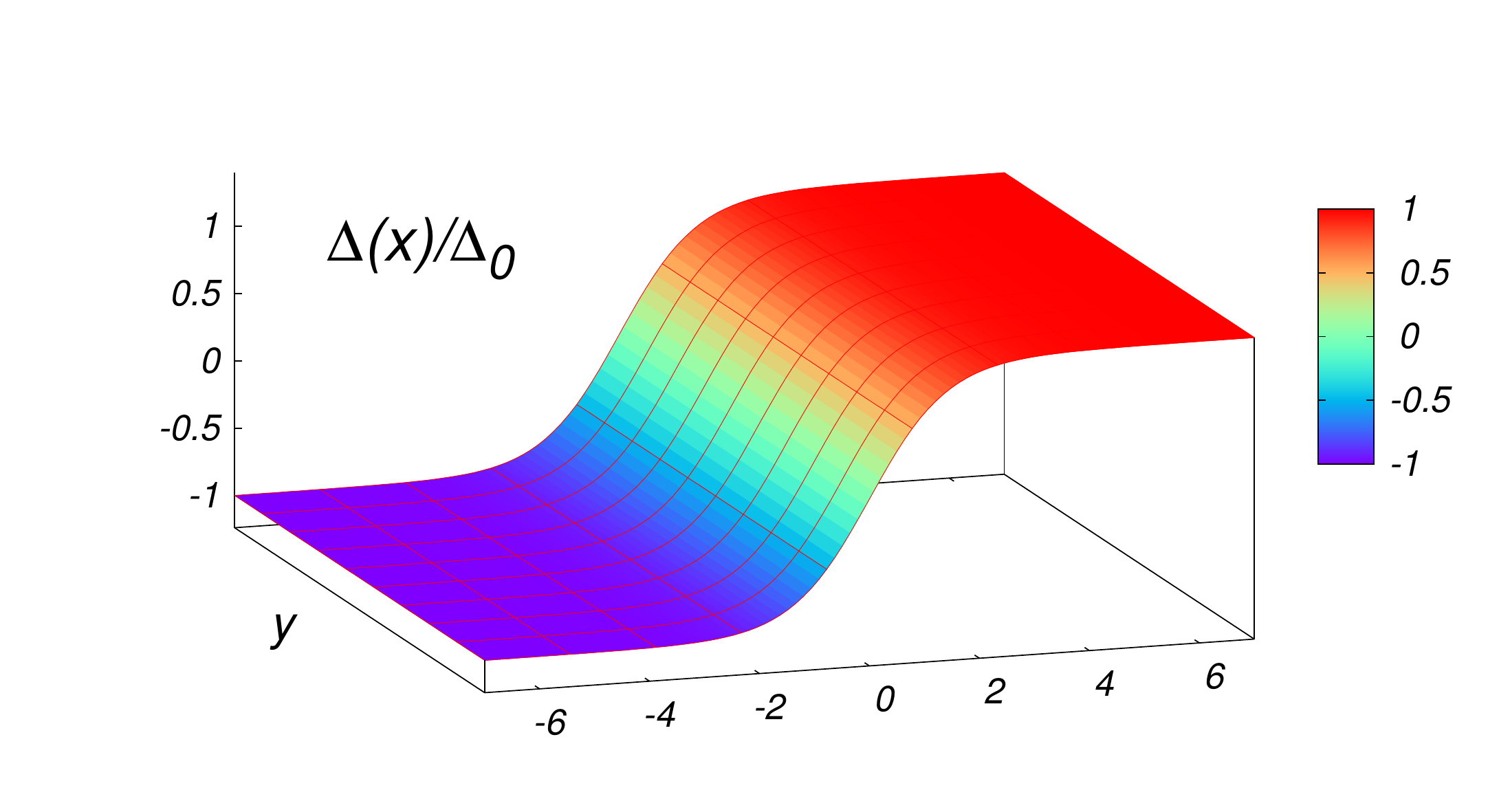}\\
\vspace*{-1cm}
\includegraphics[height = 0.5\linewidth,width = 1.0\linewidth]{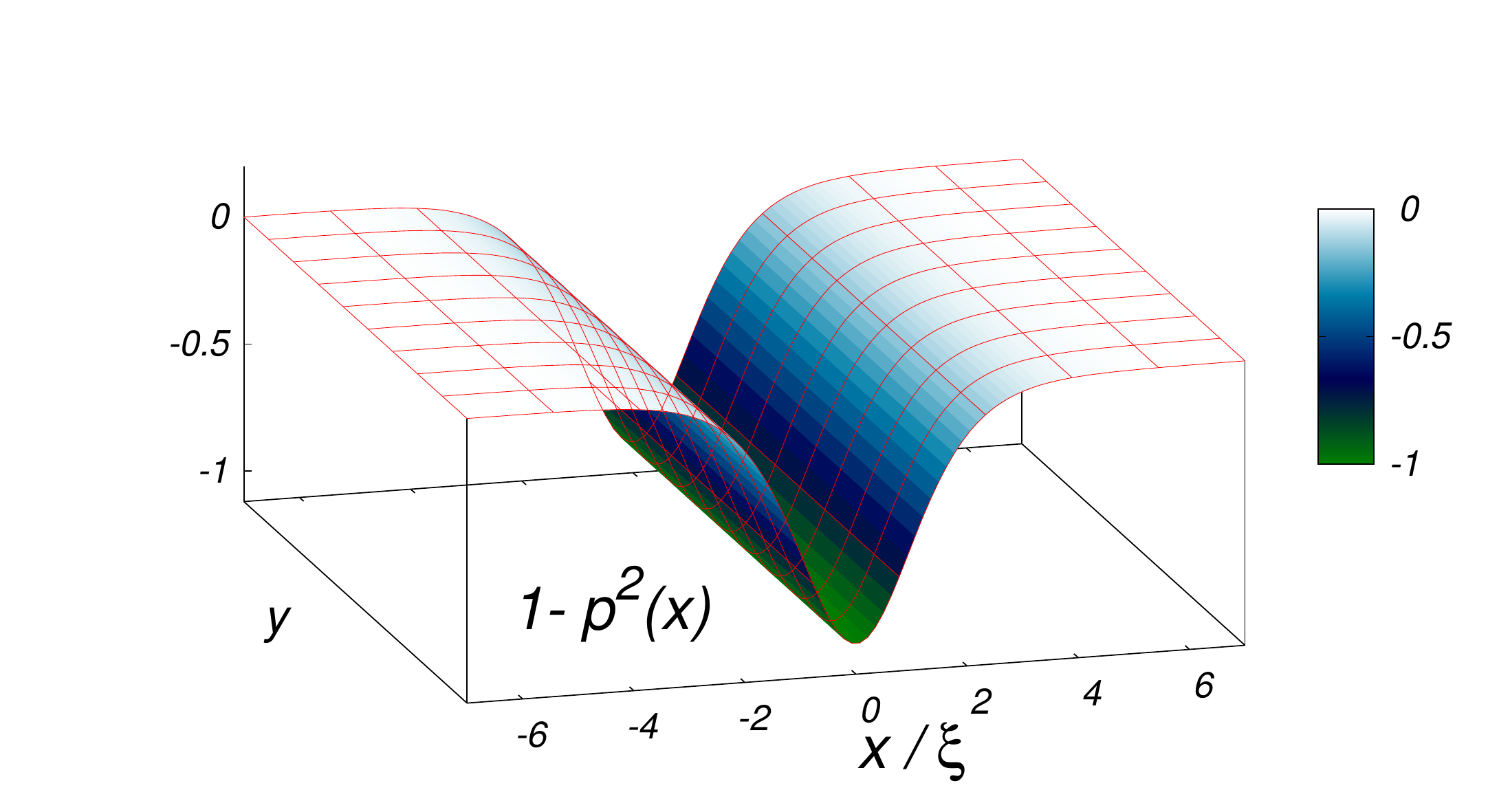}
\caption{ \label{fig:model}
(Color online) 
Domain wall with profile 
$ \Delta(x)/\Delta_0 \equiv p(x) = \tanh(x/\sqrt{2}\xi)$ 
separates two degenerate values of the order parameter $\Delta=\pm \Delta_0$ (top). 
The dynamics of the order parameter perturbations is described by  
Schr\"odinger equation with $1/\cosh^2 (x/\sqrt{2}\xi)$ potential well (bottom). 
}
\end{figure}
%%%%%%%%%%%%%%%%%%%%%%%%%%%%%%%%%%%%%%%%%%%%%%%%%%%%%%%%%%%%%%%%%%%%%%%%%%%%%%%%%

%our problem 
In this paper
we show that nonuniform superfluids or superconductors 
may provide a different avenue to investigate the amplitude/Higgs mode. 
We consider a general problem of a domain wall that breaks extra symmetries beside $U(1)$: 
\emph{translation and reflection}, as shown in Fig.~\ref{fig:model}. 
In the region of the domain wall additional amplitude 
modes exist below the pairbreaking edge, including one with gapless spectrum. 
While the free-standing domain wall is not likely, their evolution and dynamics 
is interesting from the point of view of 
frozen topological defects the early Universe.\cite{Kibble:1976fm,Zurek:1985ko,Rajantie:2003kq}
In superconductors, domain wall structures appear in 
Fulde-Ferrell-Larkin-Ovchinnikov states (FFLO),\cite{CasalbuoniRMP04} 
or in thin films.\cite{VorontsovAB:2007bs} 
Half-domain walls are more common and appear as OP suppression in the 
boundary regions of unconventional superconductors,\cite{ZHANG:1987vq,Nagato:1996vb} 
or when a singlet superconductor is in contact with a strong ferromagnet.\cite{Buzdin1988} 
Collective modes in unconventional superconductors with broken momentum-space 
symmetries have been studied in 
$d$-wave materials \cite{Barlas:2013di}; 
UPt$_3$ and UBe$_{13}$   \cite{Batlogg:1985vn,Golding:1985vh,Hirschfeld:1992wv}; 
Sr$_2$RuO$_{4}$ \cite{Fay:2000tv,Sauls:2015ui}.
%A relative phase mode exists in two-band materials and Josephson junctions.\cite{Leggett:1966hj,Blumberg:2007tx}
Superfluid \He\ feature many collective modes.\cite{Halperin1990,Dobbs1992,Wolfle:1999dr,VW}
In particular, several modes in \Heb\ phase\cite{SaulsMcKenzie} are  
easily detectable by ultrasound,\cite{Giannetta:1980up,*Calder:1980ui} 
and have evolved into a tool that can distinguish details 
of the pairing interactions on a few percent scale.\cite{Davis:2008he}
%
% experimental relevance
Distinct characteristics of bound collective modes can be used 
in detection of nonuniform superconducting states. 
Below we investigate both neutral superfluid and 
charged superconductor coupled to the gauge field. 

%~~~~~~~~~~~~~~~~~~~~~~~~~~~~~~~~~~~~~
\section{Model}
%~~~~~~~~~~~~~~~~~~~~~~~~~~~~~~~~~~~~~
%
We consider time-dependent Ginzburg-Landau (TDGL) Lagrangian,
where the order parameter field $\Delta(\vr,t)$ is minimally coupled to 
electromagnetic gauge field $(\Phi(\vr,t), \vA(\vr,t))$, 
\begin{align}
\begin{split}
\Lagr = -\gamma \left| \left(i \hbar \partial_t -2e \Phi\right) \Delta \right|^2 
+ \kappa \left|\left(\frac{\hbar}{i}\grad - \frac{2e}{c} \vA \right) \Delta \right|^2 
\\ 
-\alpha \left( |\Delta|^2 - \frac{1}{2\Delta_0^2}  |\Delta|^4 \right)
+ \frac{\vB^2 - \vE^2}{8\pi} \,.
\end{split}
\label{eq:model}
\end{align}
Here $\vB = \curl\vA,\; \vE = -\grad \Phi - (1/c)\partial_t \vA $ 
are magnetic and electric fields, and we put $\hbar=1$ from now on.
In the superconducting state below $T_c$ we take $\alpha>0$, 
and $\Delta_0$ is the real amplitude of uniform solution to GL equations without fields. 
In relativistic Lorentz-invariant theories 
$\gamma = \kappa$. 
This particular choice of $\Lagr$ agrees with 
microscopically derived 
equations of 
motion for the OP, %which are derived from a microscopic theory, 
which are of the wave type at low temperatures.\cite{AbrahamsTsuneto66,StoofTDGL1993}
From reference \onlinecite{StoofTDGL1993} we can extract low-$T$ phenomenological coefficients: 
$ 
\gamma = N_f/ 8 \Delta_0^2, \; 
\alpha = N_f/4 = 2 \gamma \Delta_0^2,\; 
\kappa = N_f v_f^2 / 24 \Delta_0^2 = {n}/{8 m\Delta_0^2},
%\kappa = N_f v_f^2 / 24 \Delta_0^2 = \gamma (v_f^2/3)= {n}/{8 m\Delta_0^2},
$
where $N_f$ is the density of states at the Fermi level for two spin projections, 
$v_f$ is the Fermi velocity, $n = N_f m v_f^2/3$ is the uniform electronic density. 
%One can generally write connection $\alpha = \gamma ({a}/{2}) \Delta_0^2$, where $a$ is a number; 
We define wave speed $v^2 = \kappa/\gamma = v_f^2/3$, 
and coherence length $\xi^2 = \kappa\hbar^2/\alpha = \hbar^2v^2/ 2 \Delta_0^2$. 

Model (\ref{eq:model}) is an adequate first step to 
investigate general relations between collective modes, topology and 
broken spatial symmetry. 
However, its main limitation is the lack of coupling to fermionic quasiparticles 
that would contribute to damping of collective modes. 
This is in part due to absence of first-order time derivative terms (diffusion), 
dominant near $T_c$,\cite{AbrahamsTsuneto66} 
which is also an indication of complete 
particle-hole symmetry that results in full decoupling of 
the amplitude and phase dynamics.\cite{Varma:2002jm,Pekker:2015ib} 
The domain wall region hosts a high density of Andreev bound states, 
that interact with collective modes and limit their lifetime.
One might expect that bound states' damping effects are similar to those of low-energy quasiparticles 
in uniform nodal superconductors. 
For example, in \Hea\ phase, collective modes are damped\cite{Wolfle:1976wi,*Wolfle:1977fk} 
but still detectable.\cite{Paulson:1977tb} 
It is then plausible that in 
some frequency range, depending on the availability of excitation phase space, 
the collective modes near a domain wall will not be overdamped.\cite{Heikkinen:2011tn} 
The complete treatment of dynamics of coupled order parameter modes, 
excitations and charge density
will require future fully microscopic calculation. 

In terms of the OP amplitude and phase, this model is
\bea
&&\Lagr = -\gamma \left[ (\partial_t \psi)^2 + \psi^2 (\partial_t \varphi+2e\Phi)^2 \right] 
+ \frac{\vB^2 - \vE^2}{8\pi} 
\label{eq:GL}
\\
&&+ \kappa \left[ (\grad \psi)^2 + \psi^2 \left(\grad\varphi - \frac{2e}{c} \vA \right)^2 \right]
-\alpha \left( \psi^2 - \frac{\psi^4}{2\Delta_0^2} \right) 
\nonumber
\eea
Finding extrema of the action 
$
\cS = \int d\vr \int dt \; \Lagr
$
with respect to amplitude $\psi$, field potentials $\vA$ and $\Phi$, 
gives the dynamics of the order parameter 
\bea
\gamma \pder{^2}{t^2} \psi - \kappa \nabla^2 \psi - \alpha \psi \left( 1 - \frac{\psi^2}{\Delta_0^2} \right)
-\gamma \psi (\partial_t \varphi + 2e\Phi)^2 
\nonumber \\
+ \kappa \psi \left(\grad\varphi - \frac{2e}{c} \vA \right)^2 = 0 \qquad
\label{eq:ampl}
\eea
and that of the gauge field:
\bea
&& \curl\vB - \frac{1}{c}\pder{\vE}{t} = \frac{4\pi}{c} \vj, \qquad 
\dive\vE = 4\pi \; \rho, %4e\gamma \psi^2 (\partial_t \varphi + 2e\Phi)
\label{eq:field}
\\
&& \vj= 4e\kappa \psi^2 \left(\grad\varphi - \frac{2e}{c} \vA \right) \,,
\quad  
\rho = - 4e\gamma \psi^2 (\partial_t \varphi + 2e\Phi) \,.
\nonumber
\eea
Minimization with respect to the phase of the order parameter $\varphi$ 
results in a statement of charge conservation, 
$\partial_t \rho  + \dive\vj=0$, that also follows from Eqs.~(\ref{eq:field}) 
as a consequence of the gauge symmetry. 
\cite{GreiterGauge2005}
%$
%(\varphi \to \varphi + \chi, \;
%\vA \to \vA + \frac{c}{2e}\grad\chi , \; 
%\Phi \to \Phi - \frac{1}{2e}\partial_t \chi)
%$.\cite{GreiterGauge2005}

A real-valued domain wall $\psi_0(x)$ in the absence of the fields, is a solution to 
$ - \kappa \psi'' - \alpha \psi \left( 1 - {\psi^2}/{\Delta_0^2} \right) =0$:  
\be
p(x) \equiv \frac{\psi_0(x)}{\Delta_0} =   \tanh \frac{x}{\sqrt{2} \xi} \,.
\ee
Free-standing kink extends from $-\infty < x < \infty$, Fig.~\ref{fig:model}. 
Half of the domain wall, $0\le x < \infty$, can be pinned by an interface 
with $\Delta(x=0)=0$. 

%~~~~~~~~~~~~~~~~~~~~~~~~~~~~~~~~~~~~~
\subsection{Neutral condensate}
%~~~~~~~~~~~~~~~~~~~~~~~~~~~~~~~~~~~~~
%
First, consider a neutral superconductor, $e=0$, where condensate is not coupled to the 
gauge field. 
The field equations, $\nabla^2 \vA - \partial_t^2 \vA /c^2=0$ 
give the electromagnetic wave with two transverse polarizations 
$\omega = ck, \; \vk\cdot \vA_\vk=0$, propagating with the speed of light. 
The dynamics of the order parameter perturbation around domain wall solution 
$(\psi_0(x),  \varphi_0=0)$ follows 
from (\ref{eq:model}) 
with substitution 
$\Delta(\vr,t) = \psi_0(x) + D(\vr,t)$. 
One introduces 
$D_{\pm} = [D(\vr,t) \pm D(\vr,t)^*]/2$, 
related to amplitude and phase fluctuations in linearized theory: 
$D_+(\vr,t) = \delta \psi (\vr,t)$ 
and $D_-(\vr,t) = i \psi_0(x) \delta \varphi (\vr,t)$. 
Equations for the amplitude and phase are, 
\begin{align}
\begin{split}
\frac{1}{v^2} \pder{^2}{t^2} D_+  - \nabla^2 D_+  
- \frac{3}{\xi^2}[ 1 - p^2(x) ] D_+ = -\frac{2}{\xi^2} D_+  \,
\\
\frac{1}{v^2} \pder{^2}{t^2} D_-  - \nabla^2 D_-
- \frac{1}{\xi^2}[ 1 - p^2(x) ] D_- = 0  \,
\end{split}
\label{eq:neutralCM}
\end{align}

In a uniform superconductor we put $p(x)=1$ and obtain 
an amplitude (Higgs) mode
$ \omega_+^2 = v^2 k^2 + 2v^2/\xi^2 = v^2 k^2 + 4 \Delta_0^2 $, 
with `mass' $2\Delta_0$,\cite{LittlewoodVarma81,*LittlewoodVarma82}
and the massless Bogoliubov-Anderson phase mode
$\omega_- = v k = (v_f/\sqrt{3})k$.\cite{Anderson1958}

%%%%%%%%%%%%%%%%%%%%%%%%%%%%%%%%%%%%%%%%%%%%%
%\subsection{`Waveguide-type' bound modes}
%%%%%%%%%%%%%%%%%%%%%%%%%%%%%%%%%%%%%%%%%%%%%%%%%%%%%%%%%%%%%%%%%%%%%%%%%%%%%%%%%
\begin{figure}[t]
\includegraphics[width = 0.99\linewidth]{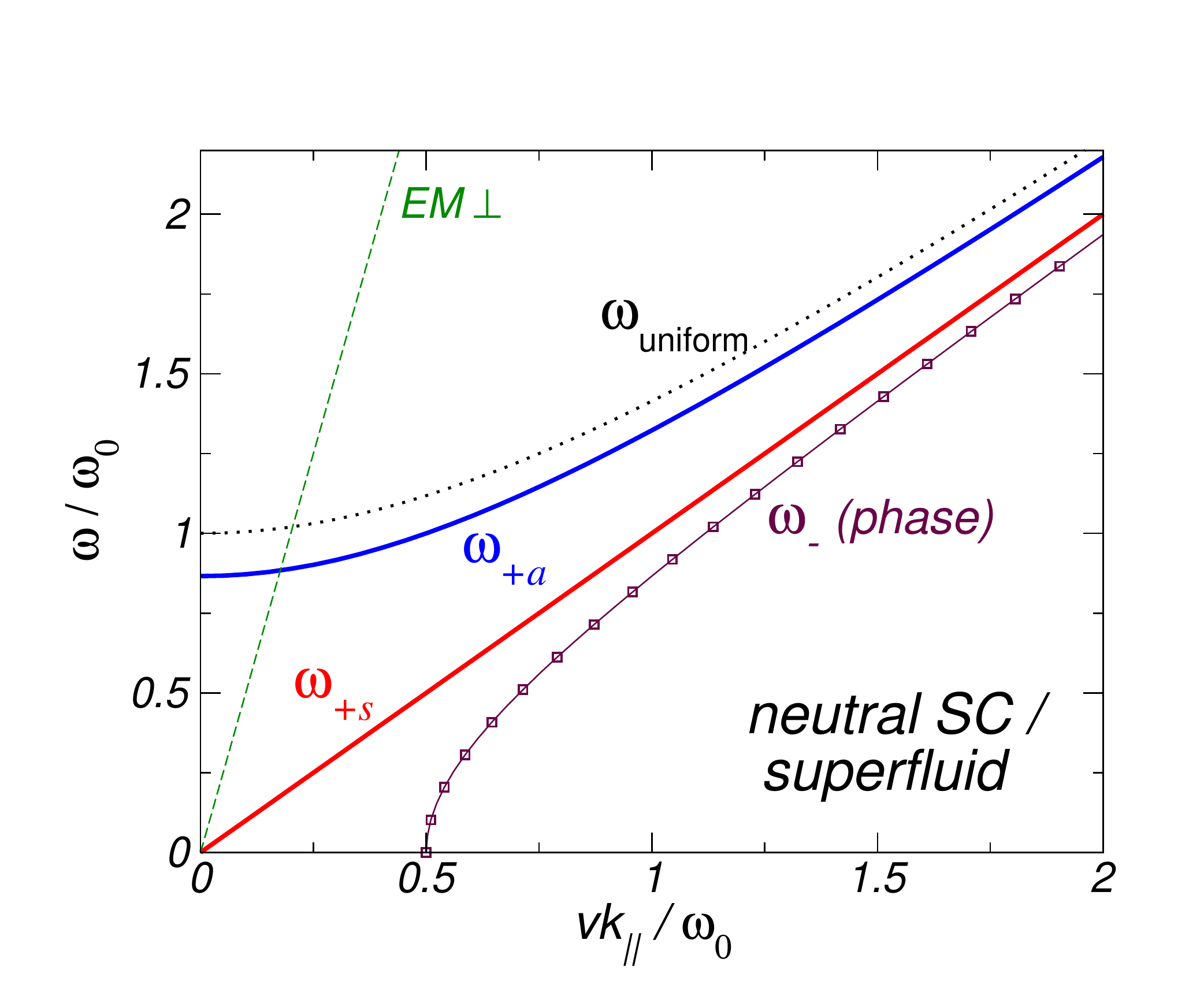}
\caption{ \label{fig:ampl}
(Color online) 
Dispersion of order parameter modes propagating along the domain wall. 
$\omega_0 = 2 \Delta_0$. 
The solid lines are the modes bound to the domain wall 
with $\omega_+ < \omega_{\sm{uniform}}$ (dotted line). 
The phase mode $\omega_-$ is unstable for long wavelengths $k<{1}/{\sqrt{2}\xi}$.
The transverse EM modes are decoupled from the order parameter dynamics. We use exaggerated $v/c=0.2$. 
}
\end{figure}
%%%%%%%%%%%%%%%%%%%%%%%%%%%%%%%%%%%%%%%%%%%%%%%%%%%%%%%%%%%%%%%%%%%%%%%%%%%%%%%%%

In the presence of a domain wall we look for collective modes that 
are localized in $x$-direction, and propagate along $y$, 
$ D_\pm(\vr,t) = D_\pm (x) e^{-i\omega t + ik_y y}$. 
For $D_\pm(x)$ pre-factors from Eq.~(\ref{eq:neutralCM}) we obtain 
\begin{align}
\begin{split}
-  D_+''
- \frac{3/\xi^2}{\cosh^2(x/\sqrt{2}\xi)} D_+ = 
\left(\frac{\omega^2}{v^2} - k_y^2 -\frac{2}{\xi^2} \right) D_+  \,,
\\
- D_-'' 
- \frac{1/\xi^2}{\cosh^2(x/\sqrt{2}\xi)} D_- = 
\left(\frac{\omega^2}{v^2} - k_y^2 \right) D_-  \,.
\end{split}
\label{eq:neutralD}
\end{align}
These equations are similar to Schr\"odinger equation for eigenstates of a particle 
in one-dimensional Eckart potential $-U_0[1-\tanh^2(x/w)] = -U_0/\cosh^2(x/w)$, 
shown in Fig.~\ref{fig:model}. 
The energies of the bound states are
$E_n = -(s-n)^2/{w^2} $ with $n < s = -1/2 +\sqrt{ 1/4 + U_0 w^2}$.\cite{LL3} 
Even/odd $n$ give symmetric/asymmetric eigenfunctions $D(-x) = \pm D(x)$. 
The OP amplitude has two bound eigenmodes ($U_0 = 3/\xi^2$, $w=\sqrt{2}\xi$, $s=2$ and $n=0,1$) 
$
{\omega_+^2}/{v^2} - k_y^2 -{2}/{\xi^2} 
= - (2-n)^2/{2\xi^2} 
$
resulting in dispersion relations
\be
\omega_{+s}^2 = v^2 k_y^2 \,, \qquad %(n=0)
\quad
\omega_{+a}^2 = v^2 k_y^2 + 3\Delta_0^2 %\qquad (n=1)
\,,
\label{eq:amplW}
\ee
shown in Fig.~\ref{fig:ampl}.
The symmetric, $n=0$, Higgs mode is massless. 
Its eigenfunction is 
$D_{+s}(x,y) \propto \exp(ik_y y) /\cosh^2(x/w)$ which can be written as 
$\tanh(x/w)\big|_x^{x+\delta x_0 \exp(ik_y y)}$ -  
a ripple of the domain wall plane. 
For $k_y=0$ it is a uniform lateral shift of entire domain wall plane without energy cost - 
consequence of spontaneously broken translational symmetry. Thus, the amplitude Higgs mode became 
a Goldstone mode, propagating along the defect with speed $v=v_f/\sqrt{3}$. 
The $n=1$ mode, in addition to translations, breaks the discrete reflection symmetry $x \to -x$ and 
corresponds to excited state of the domain wall condensate; 
it has minimal energy $\sqrt{3} \Delta_0 = \omega_0 \sqrt{3/4}$. 
Analogous results appear in extended-hadron model in field theory,\cite{Dashen1974}
and for dynamics of domain walls in structurally-unstable lattices.\cite{Wada1978}
Low-energy modes associated with dynamics 
of periodic lattice-like FFLO structures 
were explored in superconductors 
\cite{Samokhin:2010ws,*Samokhin:2011jb} and in cold atoms \cite{Rardizohovsky2009}.
%Collective modes in \Heb\ films were recently discussed in \cite{Mizushima2015Sauls}.

The phase mode ($U_0 = 1/\xi^2$, $w=\sqrt{2}\xi$, and $s=1$) 
has only one eigenvalue with $n=0$,
$ {\omega^2}/{v^2} - k_y^2  = -  (1-n)^2/{2\xi^2 }, $
and dispersion 
\be
\omega_-^2 = v^2 k_y^2 - \Delta_0^2 \,. %\qquad (n=0)
\label{eq:Wphase}
\ee
For a free-standing kink this indicates `imaginary' mass and 
instability at wave vectors $k_y < 1/\sqrt{2}\xi$, 
resulting in the decay of the domain wall, which we address later. 
For a half-kink pinned at the surface, the symmetric solutions $n=0$ are 
excluded by the boundary condition on the order parameter, $\Delta(0)=0$,  
and only the asymmetric amplitude mode propagates. 
%While the free-standing domain wall may or may not appear in otherwise uniform superconductor, 
%the half-kink can be realized in singlet superconductor-ferromagnet hybrid structures, 
%opening possibility for the detection of the amplitude modes. 

%~~~~~~~~~~~~~~~~~~~~~~~~~~~~~~~~~~~~~
\subsection{Charged superconductor}
%~~~~~~~~~~~~~~~~~~~~~~~~~~~~~~~~~~~~~
%
If $e\ne0$, the phase degree of freedom is not longer independent, 
and is absorbed into potentials $(\Phi,\vA)$. 
$
\vA \to \vA - ({c}/{2e}) \grad \varphi\,,
\Phi \to \Phi + ({1}/{2e}) \partial_t \varphi
$.
This is the unitary gauge with real order parameter, $\varphi(\vr,t)=0$. 
We assume no topological defects in the phase (vortices), that 
in this gauge would represent themselves as non-physical singularities in the 
gauge field 
(e.g. superconducting vortex $\varphi(r,\phi) \propto \phi$ gives $A_\phi \sim 1/r$ 
\cite{Tinkham,*Rajantie:2002du}).
We linearize equations (\ref{eq:ampl}-\ref{eq:field})
around zero-field domain wall solution $\psi_0(x) = \Delta_0 p(x), \Phi_0=\vA_0 =0$. 
Equation for the amplitude mode does not change from the neutral case, 
and the dispersion relations Eq.~(\ref{eq:amplW}) remain the same. 

%%%%%%%%%%%%%%%%%%%%%%%%%%%%%%%%%%%%%%%%%%%%%%%%%%%%%%%%%%%%%%%%%%%%%%%%%%%%%%%%%
\begin{figure}[t]
\includegraphics[width = 0.99\linewidth]{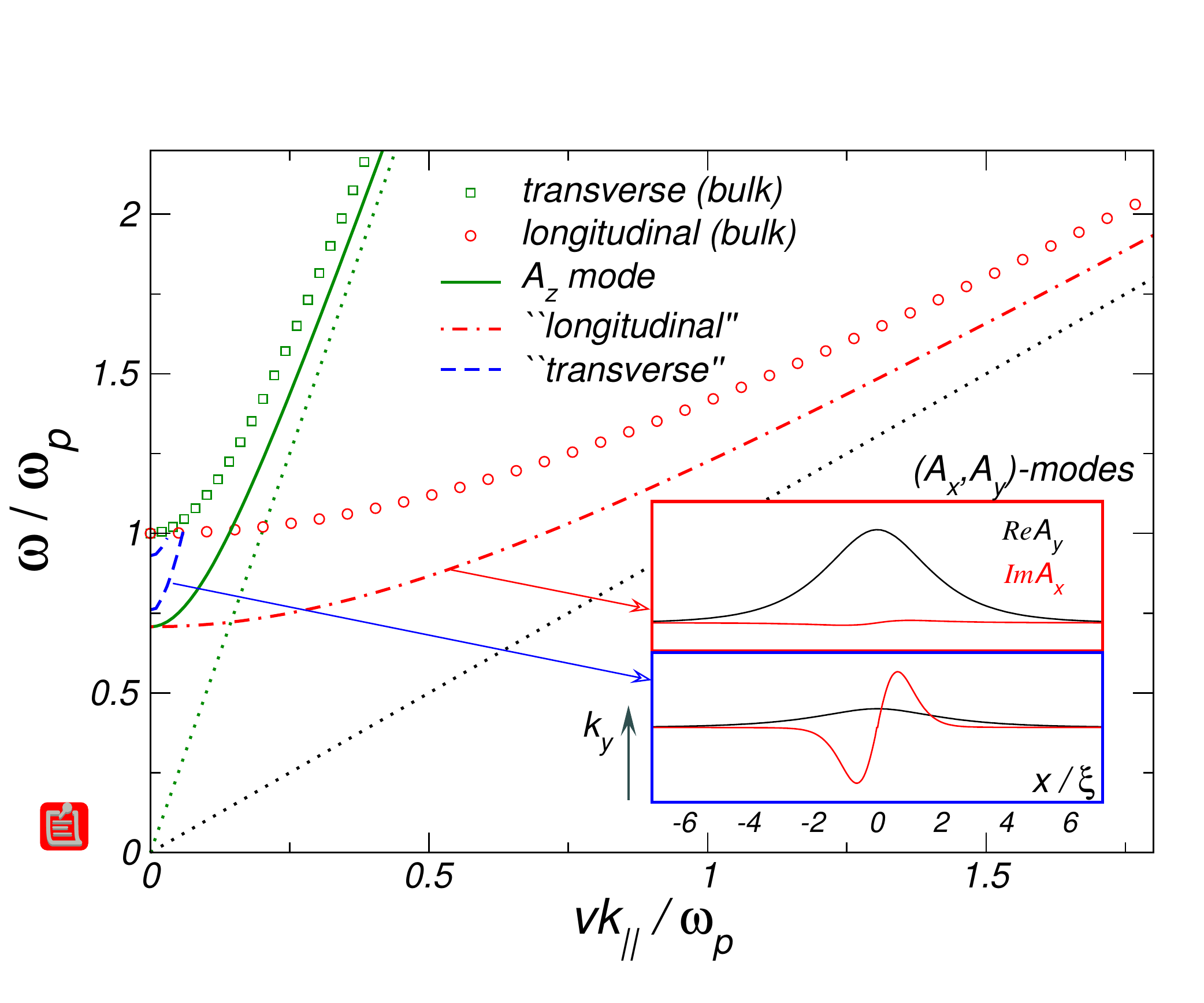}
\caption{ \label{fig:emphase}
(Color online) 
EM modes in a superconductor. 
Uniform superconductor modes are gapped with plasma frequency (open symbols). 
Modes bound to the domain wall are a transverse $A_z(x)$-mode, 
and ($A_x(x),A_y(x)$)-coupled modes. 
For chosen parameters, ${c}/{v}=5$, $\lambda/\xi=1$, 
examples of profiles for `longitudinal' and one of the `transverse' modes are shown in the inset. 
}
\end{figure}
%%%%%%%%%%%%%%%%%%%%%%%%%%%%%%%%%%%%%%%%%%%%%%%%%%%%%%%%%%%%%%%%%%%%%%%%%%%%%%%%%

Combining the continuity equation with Amp\`ere law in (\ref{eq:field}), we eliminate $\Phi$ 
and obtain a single equation for the vector potential: 
\bea
-\nabla^2 \vA +\frac{1}{c^2}\pder{^2\vA}{t^2} 
+ \grad \left( \mbox{div}\vA -\frac{v^2}{c^2}\frac{1}{p^2(x)}\mbox{div}[p^2(x)\vA]  \right) 
\nonumber \\ 
 -\frac{1}{\lambda^2} [1-p^2(x)] \vA 
= -\frac{1}{\lambda^2} \vA  \,.
\qquad
\label{eq:vA}
\eea
The magnetic penetration length is
$\lambda^{-2} = {32 \pi e^2 \kappa \Delta_0^2 }/{c^2} = {4\pi e^2 n}/{c^2 m } 
= {\omega_p^2}/{c^2}$, with plasma frequency $\omega_p^2 = 4\pi e^2 n/m$. 
In uniform superconductor this equation gives dispersion 
$\omega^2 = c^2 k^2 + \omega_p^2$ for two transverse ($\vk \vA = 0$) modes, 
and 
$\omega^2 = v^2 k^2 + \omega_p^2$ for longitudinal ($\vk \vA_\ell = k A_\ell$) mode 
that couples phase oscillations with motion of the electric charge. 
For bound waves propagating along the domain wall, 
$\vA(\vr, t) =  \vA(x) e^{ik_y y -i\omega t}$ 
we find several solutions. 
Transverse wave with $z$ polarization $\hat\vz A_z(x)$ 
satisfies equation similar to (\ref{eq:neutralD}), with 
Eckart potential amplitude $U_0= 1/\lambda^2$
and eigenvalues 
${\omega^2}/{c^2} -k_y^2 - {1}/{\lambda^2} = -{(s-n)^2}/{2\xi^2}
$ 
($
n < s=- 1/2+\sqrt{1/4+{2\xi^2}/{\lambda^2}}
$)
producing 
\be
\omega^2 = \omega_p^2(s,n) + c^2 k_y^2 
\;,\qquad
\ee
with lowered plasma frequency
$\omega_p^2(s,n) = \omega_p^2 \left[ 1-{\lambda^2}(s-n)^2/{2\xi^2} \right]$. 
For $\lambda \ge \xi$ there is only one bound solution $n=0$, 
while 
for $\lambda < \xi$ one has $s>1$ and multiple branches of the plasmon mode.  
Other modes satisfy coupled differential equations for $A_x(x)$ and $A_y(x)$, 
that we solve numerically. 
The dispersion relations and structure of these modes for $\lambda=\xi$ are shown in 
Fig.~\ref{fig:emphase}. 
These modes have a resemblance to the plasmon polariton modes that are bound to 
the interface regions between two different dielectrics, for example. 

We close this discussion by mentioning reflection properties of the domain wall. 
Traveling wave solution 
$
D_\pm(\vr,t) = D_\pm(x) \exp(-i\omega t)
$
to equations (\ref{eq:neutralCM}), with boundary conditions on far left/right
$$
D_\pm(-\infty ) \sim e^{ik_x x} + R_\pm e^{-ik_x x} \,,
\qquad
D_\pm(+\infty ) \sim T_\pm e^{ik_x x} \,,
$$
is known.\cite{LL3}
The transmission is determined by a combinations of $\Gamma$-functions: 
\be
T_{\pm} = \frac{\Gamma(-s_\pm-i k_x w ) \Gamma(s_\pm+1-i k_x w )}
{\Gamma(-i k_x w ) \Gamma(1-i k_x w )} \,,
\label{eq:transmit}
\ee
with $k_x^2 = (\omega^2 -\omega_0^2)/v^2$, $s_+=2$ for amplitude, and 
$k_x^2 = \omega^2/v^2$, $s_-=1$ for phase, modes. 
For integer parameter $s$ there is no reflected wave 
$R_\pm \propto 1/\Gamma(-s) =0$.\cite{Wada1978}
Similarly, for a transverse EM waves at normal incidence, $A_{y,z}$, 
%$A_{y,z}(x) \exp(-i\omega t) \perp k_x$,
Eq.~\ref{eq:vA} reduces again to one with $-(1/\lambda^2)/\cosh^2(x/w)$ potential. 
The transmission amplitude is given by (\ref{eq:transmit}) with 
$k_x = \sqrt{\omega^2 - \omega_p^2}/c$ and 
$s=-{1}/{2}+\sqrt{{1}/{4}+{\xi^2}/{\lambda^2}}$. 
For frequencies such that $k_x w \gg 1$ or $s\ll 1$, $|T_\perp|\sim 1$.
The longitudinal component, $A_{x}$, 
is entirely reflected, $T_{||}=0$ 
due to divergent term $1/p^2(x)$. 
%, and solution near $x=0$ is proportional to spherical Bessel function $A_x(x) \sim j_1(k_x x )$. 

%~~~~~~~~~~~~~~~~~~~~~~~~~~~~~~~~~~~~~
\subsection{Topology connection}
%~~~~~~~~~~~~~~~~~~~~~~~~~~~~~~~~~~~~~
%
%%%%%%%%%%%%%%%%%%%%%%%%%%%%%%%%%%%%%%%%%%%%%%%%%%%%%%%%%%%%%%%%%%%%%%%%%%%%%%%%%
\begin{figure}[t]
\includegraphics[width = 0.99\linewidth]{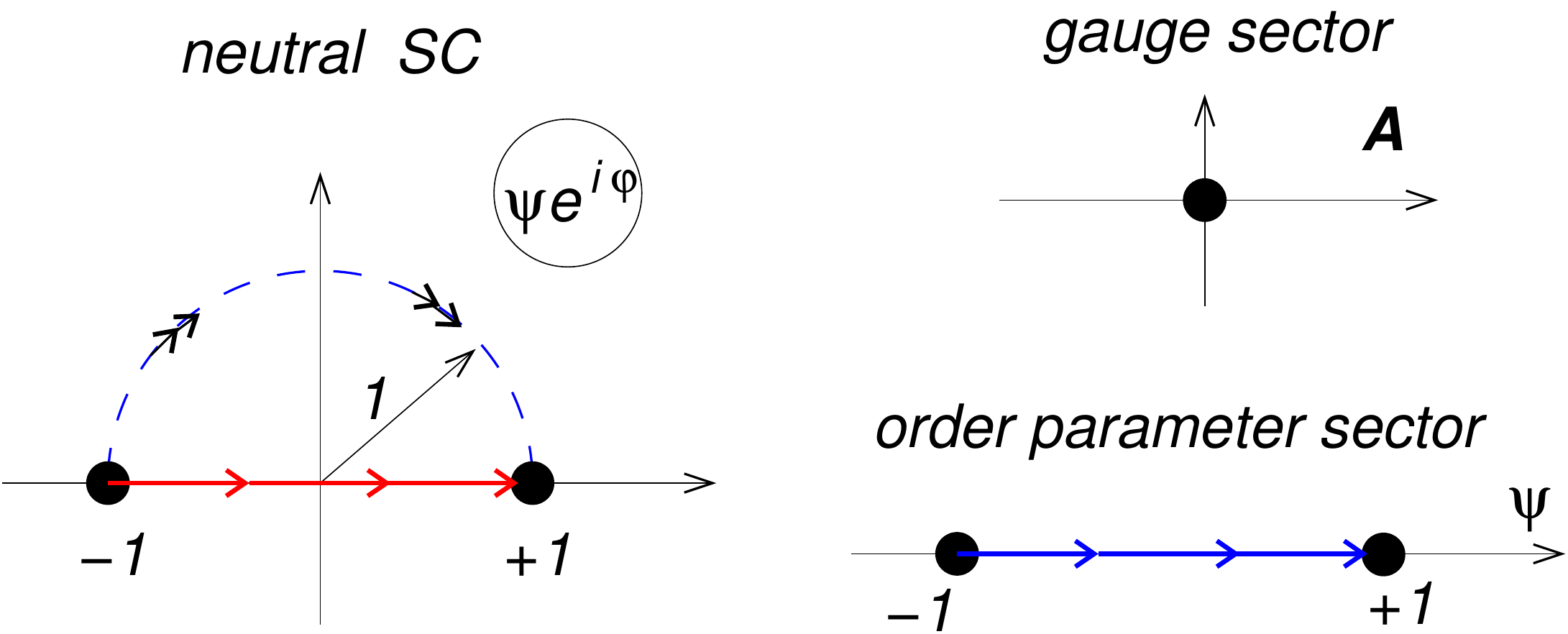}
\caption{ \label{fig:topology}
(Color online) 
In neutral condensate the real-valued domain wall (red line) is unstable 
with respect to deformations towards phase texture (dashed semicircle), 
that can be continuously deformed into a trivial uniform configuration. 
After coupling to the EM potentials, the field/phase sector separates from the amplitude sector, 
and the degeneracy space of the real OP amplitude becomes disconnected ($\pm 1$), 
stabilizing the real-valued kink.
}
\end{figure}
%%%%%%%%%%%%%%%%%%%%%%%%%%%%%%%%%%%%%%%%%%%%%%%%%%%%%%%%%%%%%%%%%%%%%%%%%%%%%%%%%

Finally, we interpret the collective mode frequencies 
in terms of topological properties of the order parameter space 
and stability of the domain wall. 
The $\omega^2_- < 0$ frequency of the imaginary component (\ref{eq:Wphase}) 
in neutral superfluid 
indicates that the real-valued domain wall is not stable. % topologically 
Indeed, the kink has energy  
%$\Delta(x)/\Delta_0 =\tanh(x/\sqrt{2}\xi) = -1 \to +1$ 
$ \alpha (4\sqrt{2}/3) \Delta_0^2 \xi$ over the uniform configuration; 
it is represented by the red line
on the left of Fig.~\ref{fig:topology}. 
An alternative solution to a hard domain wall is a long-wavelength `soft texture' 
of phase variation 
$\Delta(x)/\Delta_0 = e^{i\varphi(x)}$, $\varphi = \pi \to 0$ along 
the connected $U(1)$ degeneracy manifold, denoted by the dashed semi-circle. 
This configuration has the energy of trivial uniform state, and can be continuously deformed into one, 
due to gapless nature of the phase fluctuations.\cite{Kibble:1976fm} 
$ | \Im \omega_-|$ gives the decay rate of the hard domain wall towards the topologically trivial texture. 
In a charged superconductor the phase degree of freedom is absorbed into the gauge field sector, 
gapped with plasma frequency.
The manifold of the degenerate states of real order parameter becomes disconnected, 
containing just two points $\pm \Delta_0$, which stabilizes the topological kink. 
This manifold has $\mathbb{Z}_2$ symmetry: 
kink and anti-kink are unstable and will continuously deform into lower energy uniform 
configuration.\cite{kinks}
This also follows from 
the Schr\"odinger equation (\ref{eq:neutralCM}) 
for $D_+$ with two potential wells separated by $L$. 
In WKB solution the zero-frequency mode $\omega_+^2(k_y=0)=0$ is split, and one of the  
frequencies becomes imaginary: $\omega^2 \sim -\exp(-L/\xi)$, 
signifying instability of the double domain wall configuration. 
%One may also conclude that since this splitting is exponentially small, 
%in order to have an instability for arbitrary $L$, one has to have a massless 
%mode at a $\mathbb{Z}_2$ domain wall. 

\section{CONCLUSIONS}

In summary, a region of strongly varying condensate, such as 
a domain wall or a pairbreaking interface, 
hosts additional bound collective modes of the order parameter. 
For a single-component complex order parameter we find two additional amplitude modes 
below the bulk pairbreaking edge $2\Delta$. 
One mode lies at $1.73\Delta$, 
and the other has zero excitation mass, due to broken translational symmetry, Fig.~\ref{fig:ampl}. 
%The zero excitation gap is a requirement of $\mathbb{Z}_2$ topology of the 
%superconducting condensate in the presence of gauge field. 
The nonuniform region supports extra bound gauge field modes as well, Fig.~\ref{fig:emphase}.
Domain wall completely reflects the longitudinal component of the field and is transparent 
to others; perfectly transmistting bulk amplitude modes. 

\emph{Acknowledgements.}
This research was done with NSF support through grant DMR-0954342. 
ABV would like to acknowledge discussions with Jim Sauls, 
and hospitality of the Aspen Center for Physics, 
%supported by National Science Foundation grant PHY-1066293, 
where this work was first conceived. 

   %~~~~~~~~~~~~~~~~~~~~~~~~~~~~~~~~~~~~~~~~~~~~~~~~~~~~~~~~~~~~~~~~~~~~~~~~~~~~~~~
    %~~~~~~~~~~~~~~~~~~~~~~~~~~~~~~~~~~~~~~~~~~~~~~~~~~~~~~~~~~~~~~~~~~~~~~~~~~~~~~~
     %~~~~~~~~~~~~~~~~~~~~~~~~~~~~~~~~~~~~~~~~~~~~~~~~~~~~~~~~~~~~~~~~~~~~~~~~~~~~~~~
      %~~~~~~~~~~~~~~~~~~~~~~~~~~~~~~~~~~~~~~~~~~~~~~~~~~~~~~~~~~~~~~~~~~~~~~~~~~~~~~~

\bibliographystyle{apsrev4-1}
\bibliography{bibColMod}

%~~~~~~~~~~~~~~~~~~~~~~~~~~~~~~~~~~~~~~~~~~~~~~~~~~~~~~~~~~~~~~~~~~~~~~~~~~~~~~~%
\end{document}